# Superconductivity in porous MgB$_2$

V. Grinenko[a], E.P. Krasnoperov[a,*], V.A. Stoliarov[a], A.A. Bush[b] and B.P. Mikhajlov[c]

[a]*RRC Kurchatov Institute, 123182 Moscow, Russia;*
[b]*MIREA, 119454, Moscow, Russia;*
[c]*IMET, 119991, Moscow, Russia*

**Abstract**.

Porous magnesium diboride samples have been prepared by the heat treatment of a pressed mixture of Mg and MgB$_2$ powders. It was found that linked superconducting structure is formed down to the minimum normalized density $\gamma_c = d/d_o \cong 0.16$ (percolation threshold), where d is the density of MgB$_2$ averaged over the sample, $d_o = 2.62$ g/cm$^3$ is the X-ray density. Lattice parameters and critical temperature of the porous sample decrease with increasing porosity (decreasing γ) and $T_{c2} \cong 32$K is minimal at $\gamma_c$. The grain boundaries in the porous samples are transparent for the current and $J_c \sim 3 \cdot 10^5$ A/cm$^2$ in self field at T= 20K in the samples with $\gamma \sim 0.24$.

## 1. Introduction

The weak grain-boundary links limit substantially the critical current in ceramic high-temperature superconductors [1]. In the magnesium diboride, the inter-grain resistance is also observed after machining as consequence of the high hardness of MgB$_2$. This reduces the critical current of wires and tapes made using the powder in tube method [2, 3] in comparison with the bulk samples synthesized by hot isostatic pressing [4, 5]. It has been found that a magnesium excess (about 5 %) in an initial MgB$_2$ powder allows one to increase the critical current density of a wire by more than an order of magnitude. At T= 4.2K, the current reaches $J_c \cong 5 \cdot 10^5$ A/cm$^2$ [6]. It was noted that machining causes the formation numerous micro cracks and defects that disappear after annealing owing to recrystallization processes promoted by the Mg excess. The method of magnetic compression and the subsequent annealing allows one to obtain a high-density MgB$_2$ rod with the 860 A current in a core of 2.3 mm diameter at T= 4.2K [7]. Scanning tunnel microscopy [8] of polycrystalline bulk samples has revealed the presence of a normal inter-grain layer in the form of an amorphous region extending over 5 to 20 nm. However, according to the authors [8] the critical current was very high. On the other hand an applied small dc field (a few Gauss) in addition to an ac field leads to the generation of even harmonics, which are also an indication of the presence of some weakly linked grain boundaries in the MgB$_2$ polycrystalline bulk sample [9].

It is known that MgB$_2$ is formed by a solid-state reaction. The investigation of the crystal merging physics and properties of the interface between crystals obtained during solid-state reaction is of significant interest. In this work, porous MgB$_2$ samples down to minimal average density were prepared. Formation of the new phase and inter-grain superconductivity was investigated depending on the average density (porosity) of samples. The temperature dependence of the magnetization and resistance measured in different magnetic field, X-ray diffraction analysis and scanning electron microscopy (SEM) data are reported.

## 2. Samples and methods

Porous samples were prepared by heat treatment of MgB$_2$ and Mg mixed powders. As starting materials, we used single-phase (in terms of X-ray diffraction) MgB$_2$ (99.6 % purity) having a crystal size of less than 10 µm and Mg (99.9 % purity) having a particle size of less than 100 µm. The initial weight proportion $MgB_2$:Mg was varied from pure magnesium diboride up to a ratio of 1:7. Powders were mixed carefully and pressed into tablets of 5.1 mm diameter and about 1.5 mm in height using the 3-T press. The tablets were placed in a stainless steel container in the form of a tube which was closed by a steel stopper and annealed at temperature of 900°C for 1 – 5 hours in a He atmosphere at a pressure of 1.5-1.7 bar.

--------------------

\* *kep@isssph.kiae.ru*

The content of magnesium diboride in sample is characterized by the average density $d=M/\Omega$, where M is the $MgB_2$ mass in the pellet and $\Omega$ is the final volume of the pellet. During the heat treatment at T= 900C boron practically does not evaporate, and its quantity in the sample does not change. In case of absent $MgB_4$ phase we use the normalized average density of the magnesium diboride $\gamma = d/d_o$, where $d_o=2.62$ g/cm$^3$ is the X-ray $MgB_2$ density. Therefore $\gamma$ characterizes the average density of the magnesium diboride, i.e. the superconducting component. The $\gamma$ determination error was less than 10 %. Since the sample container was not closed hermetically, added magnesium could escape. We could control quantity of remaining magnesium in the samples by choosing the annealing time and by changing the volume of the container. In this way, porous $MgB_2$ samples with a different content of magnesium (from 20% excess to pure $MgB_2$) have been prepared. The density $\gamma$ of the annealed samples was in the range from 0.16 to 0.7.

Average diamagnetic susceptibility was measured in various fields after zero field cooling (ZFC) using a Hall sensor having area 0.5x0.5mm$^2$. The gap between pellet and sensor was about 0.7 mm. Magnetic field was directed along a perpendicular to the flat faces of the pellet. As a first step magnetic field –H (without sample) was measured. Then the local magnetic induction – B (near the pellet) was measured. Average susceptibility was calculated as $<\chi>= (B(T)-H)/4\pi H$. A standard dense pallet ($\gamma=0.94$) [7] was used to calibrate the $<\chi>$. In such samples local induction is zero (B=0) at low temperature and fields, but observed value $4\pi<\chi(T)>\approx 0.75$ was less than 1. This is explained by geometry of experiment (demagnetization factor D=0.5 and gap between pellet and Hall sensor). The error in the measured $<\chi>$ did not exceed 10%. Beside magnetic properties, resistance transitions to the superconducting state were studied. The resistance was measured using a four-probe method on a rectangular bar 5x2x1.5 mm$^3$ in size that was cut from a tablet.

### 3. Results and discussion

The fracture of annealed tablets was investigated by SEM. Photographs of the porous sample with $\gamma \cong 0.24$ are shown in Fig.1. Elongated cavities can be seen (on the left) that were formed after magnesium evaporation. The granules have size from one up to tens of microns. The X-ray diffraction analysis in this sample has shown the presence of about 5 % Mg and less than 2% of MgO. The weight of this sample is ~ 5 % more, than initial weight of $MgB_2$. On the right a fragment of the sample with the same $\gamma$ under high magnification is shown. It can be seen that size of the smallest fragments is of the order of 1 µm.

Diamagnetic susceptibilities $-4\pi<\chi(T)>$ of the samples versus temperature before and after heat treatment are shown in Fig. 2. They were obtained in the field H= 3 Oe (ZFC). Initial tablets ($\gamma \cong 0.28$ extruded in Fig. 2) have a small susceptibilities which goes to zero at a temperature equal to the critical temperature $T_{co} \approx 38K$ corresponding to a bulk sample [5, 6, 7]. The weak diamagnetism in extruded sample is explained by the fact that $MgB_2$ particles in a magnesium matrix have no superconducting interface. After annealing porous samples acquire a large diamagnetic susceptibility at low temperatures, and $<\chi(T)>$ is the same as a standard dense ($\gamma=0.94$) sample has. Large value of diamagnetic susceptibility suggests that the external magnetic field is screened completely by inter-grain surface currents. Resistance measurements (see below) have shown that the linked superconducting structure exists not only on the surface but also throughout the whole volume of the porous samples.

The value and behavior of $<\chi(T)>$ depends on the annealing time. However, after a 2-h anneal at T= 900C for the samples with initial density $\gamma> 0.16$ the susceptibility was found to be virtually unchanged and looks like it is shown in Fig. 2. Moreover, the composition of samples can be changed from 20 % Mg excess up to 15 % content of non-superconducting $MgB_4$ phase after a 4-h anneal. However, $<\chi(T)>$ dependence is identical within the accuracy of measurements. As seen in Fig.2 the dense samples (with $\gamma> 0.7$) exhibit a single superconducting transition. The susceptibility of the dense sample with $\gamma= 0.94$ [7] is labeled by asterisks. With increasing porosity (decreasing $\gamma$), two superconducting transitions are observed. The first (at $T_{co}$) is attributed to the initial $MgB_2$ phase, and the second (with a higher susceptibility) characterizes the disappearance of the macroscopic currents. We believe that the second superconducting transition (at $T_{c2}$) appears due to the new phase which arises after annealing, between initial crystallites of $MgB_2$.

Comparison of the initial and annealed tablets (curve 3 and extruded $\gamma \cong 0.28$ on Fig. 2) in a range of $T_{c2}< T< T_{co}$, allows us to see that, with the appearance of diamagnetism in the porous medium, the susceptibility of the initial $MgB_2$ phase decreases, whereas their critical temperature $T_{co}$ remains unchanged. The reduction of $<\chi>$ magnitude is caused by the reduction of the volume of the initial crystallites phase due to formation of a new phase.

Annealing of tablets with an initial density of $\gamma< 0.16$ leads to a reduction of its size both in height and in diameter and $\gamma$ increased accordingly. Left curve in Fig.2 shows dependence of the susceptibility versus temperature for the sample with initial density $\gamma \cong 0.11$. After heat treatment the density has increased up to $\gamma \cong 0.16$. Critical temperature $T_{c2}$ of this sample coincides with the $T_c$ of a sample having the same initial density $\gamma \cong 0.16$ (curve 4). It is seen, that the magnitude of the susceptibility of this tablet is less noticeably (curve 5). This is due to the reduction of tablet size after treatment, and also because the density of the sample is near to percolation threshold $\gamma_c$. Linked superconducting structure does not exist for sample density below $\gamma_c$. In such samples the superconducting linked structure arises only when the density reaches a critical value. The critical density $\gamma_c \cong 0.16$ is explained in terms of a percolation theory [10]. In case of $MgB_2$ the superconducting linked structure throughout the whole volume of the sample is observed at $\gamma \geq \gamma_c$. Supposing the grains are distributed randomly and independently of each other, $\gamma$ is the probability that the site in the porous media is occupied by a superconducting grain and $1- \gamma$ is probability that the site is empty. In this case the theoretically predicted lower limit for percolation threshold in 3-D space is $\gamma_c= 0.162$ [11]. This value coincides with minimum density of porous $MgB_2$.

The highly porous $MgB_2$ samples (near to $\gamma_c$) are very brittle. For resistance measurements mechanically strong samples are needed. In this case we used samples with a magnesium excess of about 20% (by weight). The temperature dependence of the normalized resistance $\rho(T)= R(T)/R(39K)$ measured in different fields for the sample with $\gamma \cong 0.24$ are shown in Fig. 3. For comparison, $-4\pi<\chi(T)>$ dependence is presented. The $\rho(T)$ as well as $<\chi(T)>$ shows two steps and characterizes two phases, i.e. initial granules and the new phase. It is obvious that, after disappearance of the inter-granular superconducting currents (at $T > T_{c2}$), the resistance should appear. It is seen from fig. 3, that in a 3Oe field obtained from magnetic measurements $T_{c2}$ is noticeably smaller, than $T_{c2}(0) \cong 34K$ obtained from resistance data (H=0). In order to determine $T_{c2}$ in zero field, susceptibility was measured in magnetic field range 3 – 200Oe and results were extrapolated using square approximation $H \propto (1-T_{c2}/T^*_{c2}(0))^2$. The instance of this dependence is shown in inset in fig. 4. The $T^*_{c2}(0)$ obtained from such extrapolation coincides with $T_{c2}(0)$ at which resistance become zero.

The main characteristics of samples are composed in Table 1.

TABLE I. Sample list with average density - γ, critical temperatures - $T_{co}$, $T_{c2}(0)$ and lattice parameters **a** and **c**. The next designations are used: pm- pressed mixture of $MgB_2$ and Mg before annealing with different- γ; sd - standard dense $MgB_2$ sample [7]; ht - samples after two hour heat treatment.

| Samples | γ | $T_{co}$, K | $T_{c2}(0)$, K | a, A | c, A |
|---|---|---|---|---|---|
| pm | - | 38.2 | 38.2 | 3.083(1) | 3.520(1) |
| sd | 0.94 | 38.0 | 38.0 | - | - |
| ht | 0.71 | 37.8 | 37.8 | 3.079 | 3.520 |
| ht | 0.46 | 38.1 | 36.4 | - | - |
| ht | 0.37 | 38.1 | 36.1 | - | - |
| ht | 0.31 | 38.0 | 35.0 | - | - |
| ht | 0.24 | 38.0 | 34.0 | 3.077 | 3.520 |
| ht | 0.16 | 37.9 | 32.0 | 3.074 | 3.517 |

The dependence of the relative critical temperature of the new phase $T_{c2}(0)/T_{co}$ versus density γ is shown in Fig. 4. It can be seen that the critical temperature decreases when the density decreases and one has minimal $T_{c2} \cong 32K$ at the percolation threshold. X-ray analysis has shown that in a new phase with decreasing γ the unit cell volume become smaller. As seen in Tab.1 the inter-plane distance **c** doesn't change practically in compare with **a** - distance between atoms in the plane. It is necessary to note that **a** and **c** obtained from mixture of the $MgB_2$ crystallites and a new phase are the averaged values. Therefore actual lattice parameters of a new phase are less than those shown in Tab.1. It is known that partial substitution of the Mg or B sites by other atoms reduce the $T_c$ and the unit cell volume [12, 13]. To obtain the $T_c \approx 30K$ one needs to replace the 10-15% atoms. In our samples very pure components were used so the admixture did not exceed 0.1%.

Probably the oxygen inside the $MgB_2$ lattice can deteriorate the superconducting properties. The existence and precipitation of oxygen in $MgB_2$ have been observed experimentally [14]. However, many $MgB_2$ samples containing MgO precipitates, which indicate that an oxygen source is available during growth, did not cause a significant decrease in $T_c$ [15]. But in films it was observed an apparent $T_c$ reduction [16]. The authors believe that oxygen alloying lead to the expansion of **c**-lattice parameter. In our samples with decreasing $T_c$ **a**-lattice parameter become smaller but **c** practicality doesn't change. We varied the purity of the He atmosphere for preparing porous samples, which has led to the formation of magnesium oxide phase from 2 % up to 10-15% (wt) but no changes were observed in superconducting transition. Also existence of magnesium vacancies in samples is unlikely because $T_c$ doesn't depend on remained magnesium excess after annealing. Therefore the formations of boron vacancies during the annealing are very probable and we suppose that defects in boron sublattice lead to reduction of lattice parameters and critical temperature.

A double step transition in weakly connected structures is interpreted often as arising from intra- and inter-granular transition [17, 18]. In this case grains have critical fields and currents on the order of magnitude larger then inter-granular weak links have. In [19] resistive transition in HTS had been investigated. Authors had shown that superconducting transition has two distinct sections, a steep part associated with the onset of superconductivity in the individual grains and a transition tail due to the weak links coupling the grains. With applied field the steep section remained unchanged while the tails moved considerably to lower temperatures. Magnetic measurements [18] had established also that magnetic flux penetrates into inter-granular weak links in the fields significantly lower then into grain. But in our case resistance measurements of porous samples has shown that magnetic field shifts onset of superconductivity and tails to lower temperatures approximately in the same way (see fig.3). Additionally from the magnetic measurements it was found that magnetic flux begins to penetrate into porous samples at external field of about 100Oe at 15K. This field is comparable with the first critical field $H_{c1}$ reported for pure dense bulk $MgB_2$ [20]. Additionally we estimated the average critical current density ($<J_c>$) in porous sample from the magnetization curves in terms of a Bean model [21]:

$$<J_c> \approx \frac{30(M^+ - M^-)}{d},$$

where $M^+$ and $M^-$ are the irreversible magnetization measured in the ascending and descending branch of the hysteresis loop, d is the diameter of the sample. The local current density ($J_{cm}$) in the porous medium is considerably higher then $<J_c>$ because the sample density is very small. Above $\gamma_c$ percolation theory [22] predicts the following expression for local current density:

$$J_{cm} = \left(\frac{\gamma - \gamma_c}{1 - \gamma_c}\right)^{-1.76} <J_c>$$

The term in front of $<J_c>$ can be interpreted as ratio of the total cross sections of the dense sample (the superconductor occupies all volume of the sample) to the cross section of the all currents paths existing at a certain γ in the porous sample. We also assume that all current paths have the same critical current density. In this way we obtained $J_{mc} \approx 3 \cdot 10^5$ A/cm$^2$ at T= 20K in self-field for the porous sample with $\gamma \cong 0.24$. This value is comparable with $J_c$ reported for the dense $MgB_2$ [4, 5, 6]. Therefore we believe that grain boundary links are transparent for the current. This agrees with the conclusion [8, 23] that the boundaries do not act as barriers to the current percolation path.

To conclude, after the $MgB_2$ annealing in the Mg atmosphere a homogeneous superconducting phase is formed between granules. The allied $MgB_2$ structure exists up to the critical density $\gamma_c \cong 0.16$, close to the percolation threshold. With decreasing average density of $MgB_2$ the critical temperature and lattice parameters of the porous samples decreased. Probably, a new phase is non- stoichiometric magnesium diboride with boron vacancies. Weak links have no visible effect on the superconducting properties of the porous samples.

**Figure captions**

Fig.1. The SEM photograph of the porous sample with $\gamma \cong 0.24$.
Fig.2. Susceptibility $-4\pi<\chi(T)>$ (after ZFC, H=3 Oe) versus temperature for the porous $MgB_2$ with the different γ.
Fig.3. Susceptibility (in left) and normalized resistance (in right) for different fields versus temperature. Sample has $\gamma \cong 0.24$. Approximating lines determine $T_{c2}$ in which superconducting currents disappear (H=3Oe) and $T_{c2}(0)$ for zero resistance (H = 0).
Fig.4. Normalized critical temperature of the new phase versus the $MgB_2$ density γ. The insert shows experimental values $T_{c2}$ at different fields and approximating curve for a sample with $\gamma \cong 0.24$.

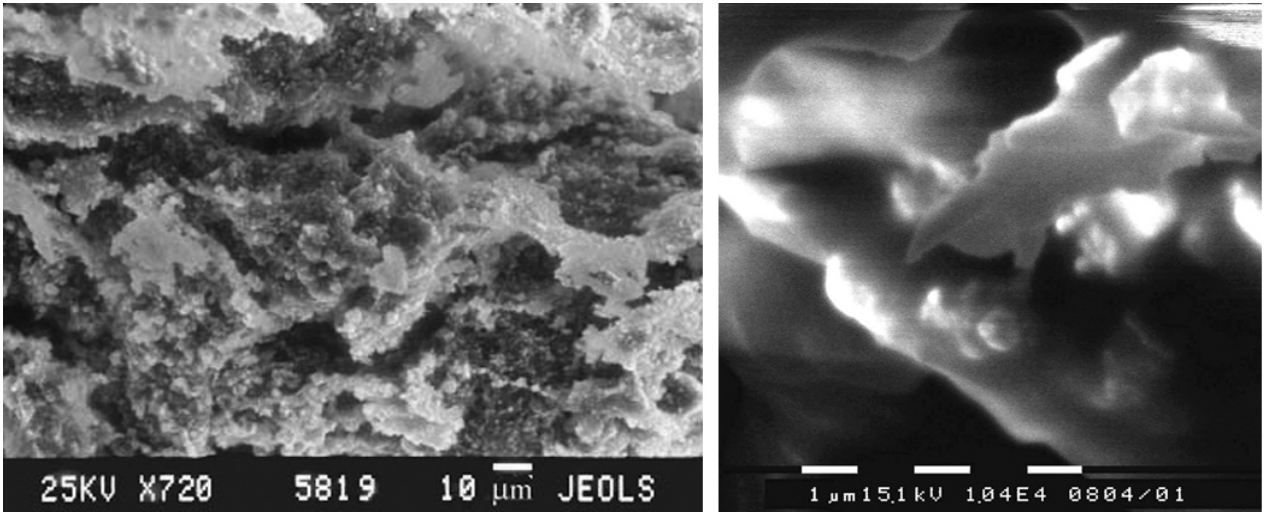

Fig. 1

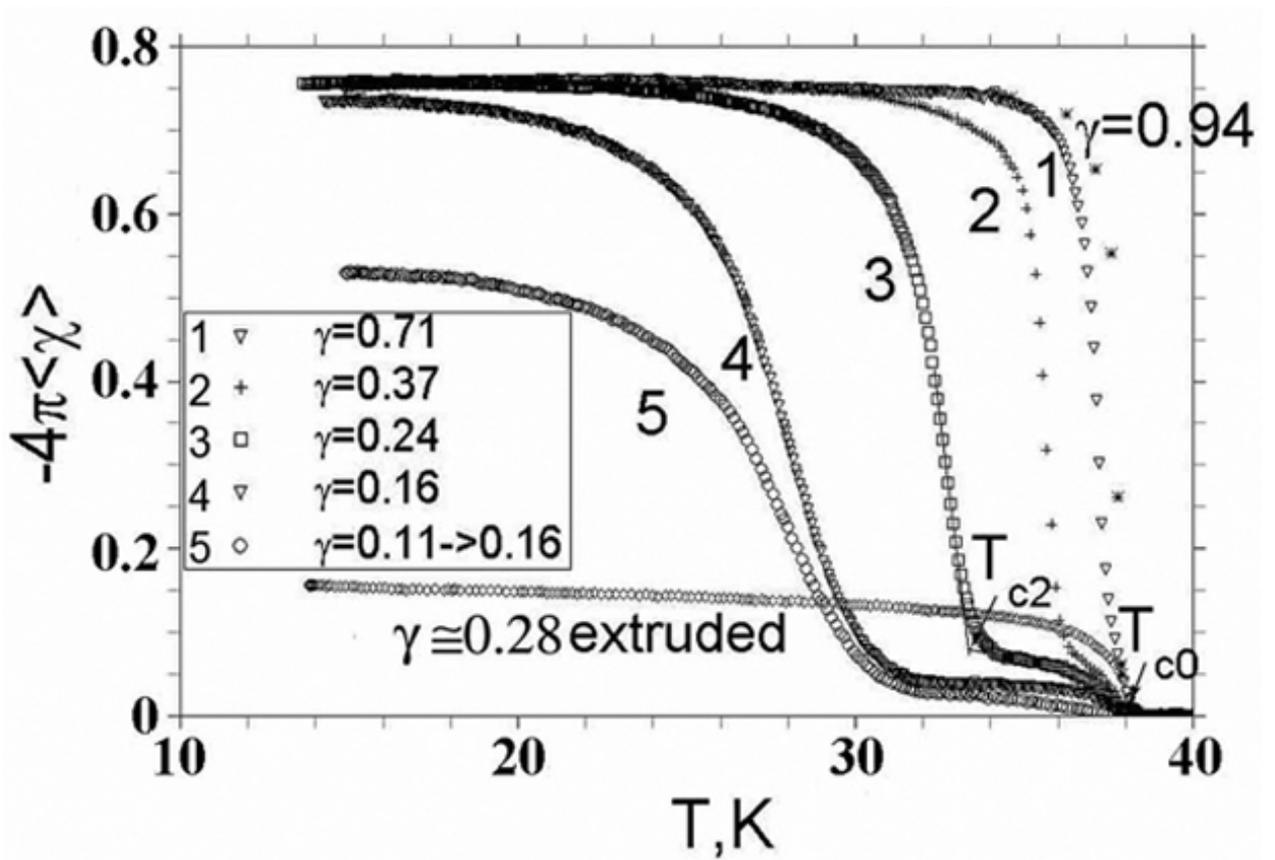

Fig. 2

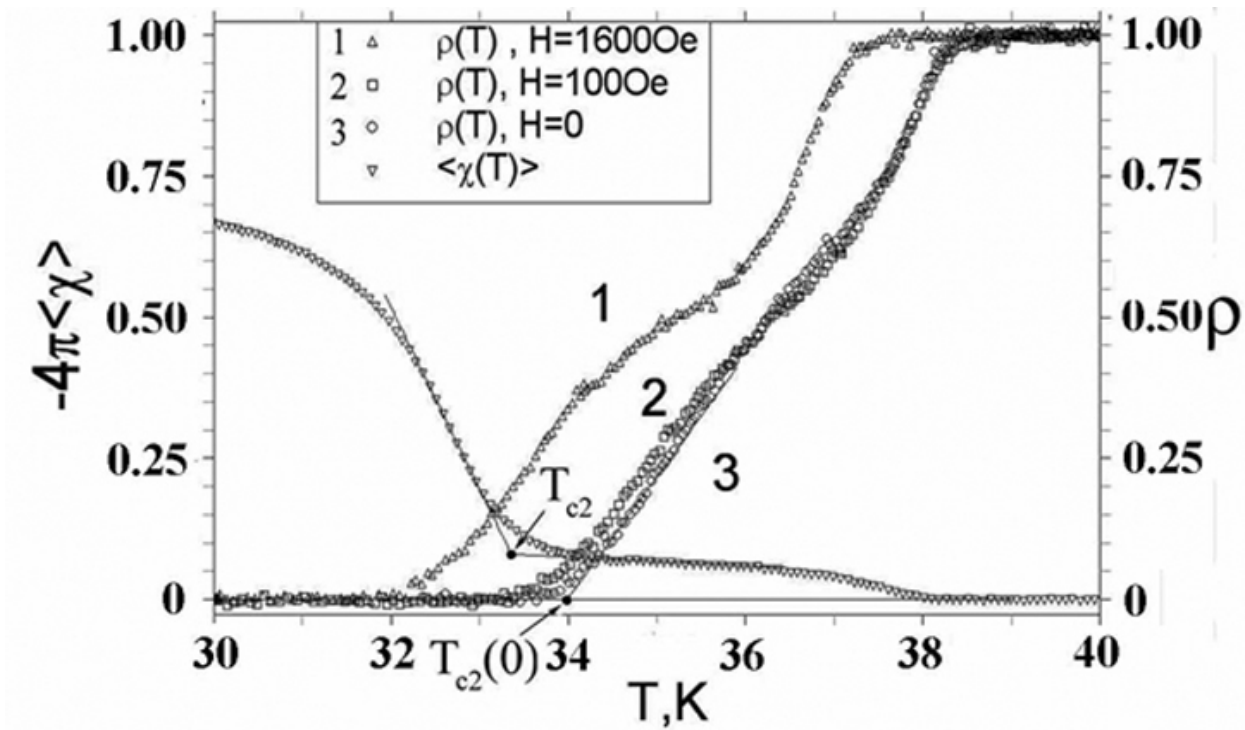

Fig. 3

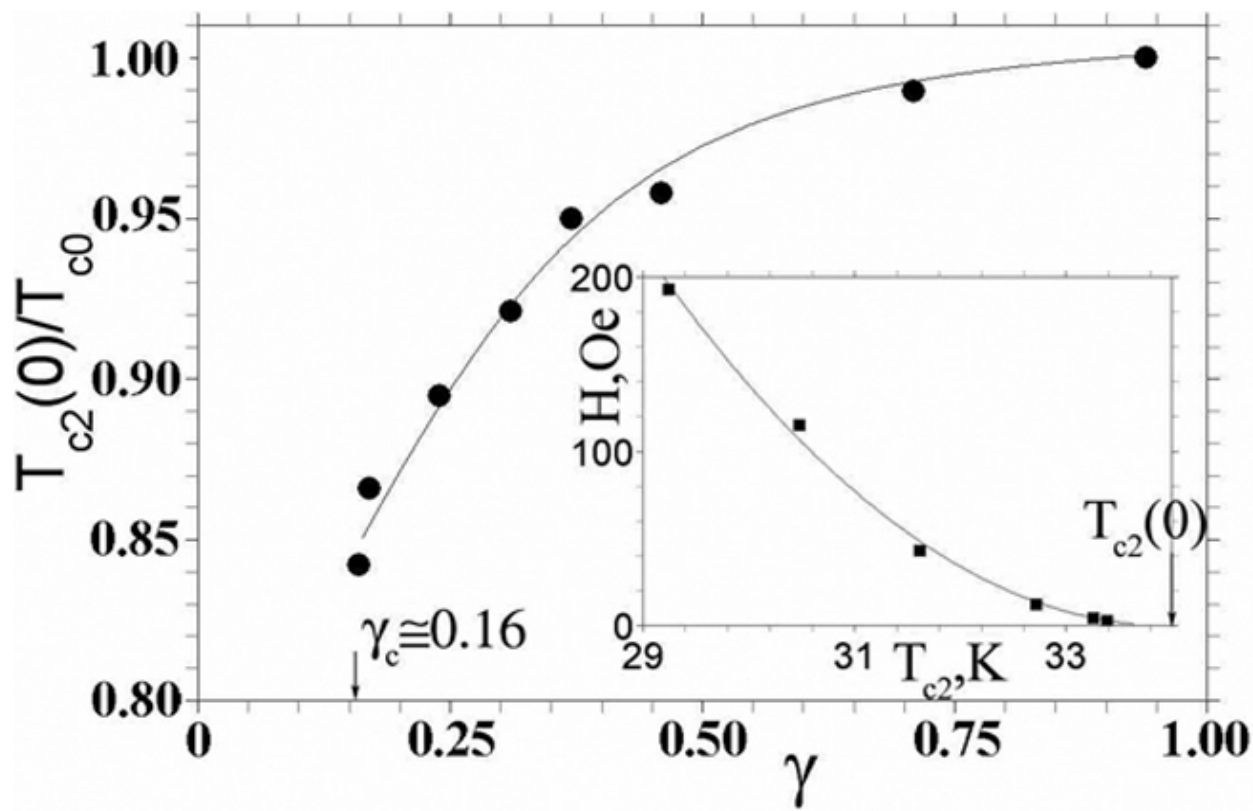

Fig. 4